\documentclass[11pt,journal,onecolumn]{IEEEtran}

\usepackage{algorithm}
\usepackage{algorithmic}
\usepackage{amsmath}
\usepackage{amsthm}
\usepackage{booktabs}
\usepackage{cite}
\usepackage{cleveref}
\usepackage{eqparbox}
\usepackage{geometry}
\usepackage{graphicx}
\usepackage{subcaption}

\newcommand{\RESULT}{\\\hspace{-1.5em}\textbf{Result: }}

\newtheorem{definition}{Definition}

\begin{document}

\title{Local Editing in LZ-End Compressed Data}

\author{
\IEEEauthorblockN{Daniel Roodt\textsuperscript{1}, Ulrich Speidel\textsuperscript{2}, Vimal Kumar\textsuperscript{1}, Ryan K. L. Ko\textsuperscript{3}}\\
\vspace{0.05in}
\IEEEauthorblockA{\small{\textsuperscript{1}University of Waikato, Hamilton, New Zealand}}\\
\IEEEauthorblockA{\small{\textsuperscript{2}University of Auckland, Auckland, New Zealand}}\\
\IEEEauthorblockA{\small{\textsuperscript{3}University of Queensland, Brisbane, Australia}}\\
\vspace{0.05in}
\IEEEauthorblockA{contact email: \emph{dkr3@students.waikato.ac.nz}}\\
}

\maketitle

\begin{abstract}
	This paper presents an algorithm for the modification of data compressed using LZ-End, a derivate of LZ77, without prior decompression. The algorithm's performance and the impact of the modifications on the compression ratio is evaluated. Finally, we discuss the importance of this work as a first step towards local editing in Lempel-Ziv compressed data.
\end{abstract}


%
\IEEEpeerreviewmaketitle

\section{Introduction}

The demand for compressed data storage has led to the development of methods to search in~\cite{tao2004lzw, lz77_pattern_matching} and randomly access compressed data~\cite{lzend, RA_TIME_SERIES, grossi_dynamic_random_access}. In addition, some purpose-built algorithms allow editing parts of the compressed data string~\cite{cram}. Recent innovations allow adapting any universal compressor to allow random access~\cite{tatwawadi, vatedka2019local}, and random access with local editing~\cite{vatedka2019local}. All of these have been achieved without decompressing the entire string, which processing compressed data traditionally requires.

The limitations of the algorithms of Tatwawadi \textit{et al.}~\cite{tatwawadi} and Vatedka and Tchamkerten~\cite{vatedka2019local} are that neither method applies to the parsing algorithm directly. Both algorithms involve breaking the data into a series of blocks which are compressed independently of one another by an arbitrary parsing algorithm (e.g. LZ77 or PPM). The local editing of the data is achieved by the method in which the blocks of data are separated and merged together~\cite{vatedka2019local}. Similarly, random access is achieved by splitting the data into blocks, some of which are represented as a sparse vector. This sparse vector is compressed so as to allow random access~\cite{mazumdar_local_recovery_general_sources, buhrman_bitvectors}. Although the LZ compression algorithms can be modified in this way to allow random access and edits, neither of these methods apply to the parsing algorithm directly. 

In 2010, Kreft and Navarro developed LZ-End, an adaptation of LZ77, to allow retrieval of arbitrary sections of data without global decompression~\cite{lzend}. Other algorithms, such as LZ78 and LZW~\cite{10.1007/3-540-45123-4_16, tao2004lzw}, allow searching over compressed data, but not extraction of arbitrary pieces of data. In this paper, we show how to use this same feature of the LZ-End algorithm to allow local editing of the compressed data without global decompression. This approach is the first step towards efficient local editing in LZ-compressed data. Each edit made to the compressed data results in a degradation of the compression ratio. We evaluate how different kinds of edits affect the compression ratio, and conclude by discussing the importance of this algorithm as an inital step towards locally editing general LZ data.

\section{The LZ-End Compression Algorithm}
\label{lzend_explanation}

Like LZ77, LZ-End encodes the data as a series of~\textit{phrases}, each of which consists of a pointer to a previous phrase and an ``innovation'' symbol which extends that previous phrase (see Section~\ref{lzend_explanation}). The main difference between LZ77 and LZ-End is that LZ-End restricts where each phrase may end. This allows LZ-End the retrieval of arbitrary data independently of any other part of the compressed string~\cite{lzend}.
Kreft and Navarro used their first version of LZ-End in 2010~\cite{lzend} to build a self-index~\cite{lzend_self_index}. Later variations of LZ-End allow parsing in linear time~\cite{lzend_linear_time} and compressed space~\cite{lzend_compressed_space}.

\subsection{Representation of the Compressed Data}
LZ-End builds on the LZ77 algorithm~\cite{lz77}, such that LZ-End allows decompression of an arbitrary portion of compressed data without decompressing any other part of the compressed data.
This section gives a rough overview of LZ-End, as presented in~\cite{lzend}.

\begin{definition}
	
	The LZ-End parsing of a series of symbols $\mathbf{T} = T_0 \dots T_{n-1}$ is a sequence $\mathbf{P}$ of $n'$ many phrases. If the first $q$ many phrases of $\mathbf{P}$ encode the first $i$ many symbols of $\mathbf{T}$, then the $q^{\mathit{th}}$ phrase of $\mathbf{P}$ encodes the longest prefix of $T_i \dots T_{n-1}$ which is a suffix of one of the phrases $P_0 \dots P_{q-1}$ \cite{lzend}.

\end{definition}

The compressed output of the LZ-End algorithm is a series of phrases, each represented conceptually as a tuple $\langle q, l, s\rangle$: 
\begin{itemize}
	\item $q$: a pointer to the previous phrase which the current phrase builds upon.
	\item $l$: the number of symbols encoded by the current phrase.
	\item $s$: the final symbol of the current phrase.
\end{itemize}

Each phrase encodes either:
\begin{itemize}
	\begin{samepage}
	\item a single symbol. In this case, $q$ is {\em null} and  $s$ is the symbol encoded.
	\end{samepage}
	\item a series of symbols. $q$ points to the last phrase which the current phrase extends. By making $l-1$ greater than the length of the phrase at index $q$, this pointer may reference multiple adjacent phrases, and the remaining symbols of the current phrase are derived from phrases $q-1$, $q-2$, etc. As in LZ77, $s$ extends this sequence of symbols from the previous phrases~\cite{lzend}.
\end{itemize}

Unlike LZ77, LZ-End parsing does not use a sliding window~\cite{lzend}, but parses by reading the whole string into memory and constructing a suffix array of the data~\cite{lzend}. Three arrays store the compressed data. To compress a string ${\bf T}$ of $n$ symbols ${\bf T}=T_1 T_2 \dots T_n$ into $n'$ phrases:
\begin{itemize}
	\item char[$n'$] stores $s$ in the tuple above for each phrase,
	\item source[$n'$] stores $q$ in the tuple for each phrase, and 
	\item $\mathbf{B}[n]$, where $\mathbf{B}[i] = 1$ if $T_i$ is at the end of a phrase, 0 otherwise. This array is sparse and stored as a list of indices to the non-zero elements (see~\cite{raman_indexable_dictionary}). This supports two constant-time operations:
		\begin{itemize}
			\item $\mathit{rank}_B(p)$: The number of 1's in $\mathbf{B}$ at indices up to $p$, i.e., the number of phrases preceding $T_p$.
			\item $\mathit{select}_B(p)$: The position of the $p$-th 1 in $\mathbf{B}$, i.e., the index of the last symbol in $\mathbf{T}$ encoded by the $p$-th phrase.
		\end{itemize}
		Collectively, $\mathit{rank}_B()$ and $\mathit{select}_B()$ yield the length~$l$ of each phrase.
\end{itemize}

\subsection{Notation}
When a phrase $y$ points back to a phrase $x$, we say that $y$ {\em depends on} $x$, and we call $y$ a {\em dependent} of $x$. Note that $y$ may depend on any number of symbols from $x$ and its immediate predecessors. A phrase may have any number of dependents, including none. Let $[a,b]$ denote the range of integers between $a$ and $b$ inclusive. 
When referring to a range of variables in an array, we use $T[i, j]$ to refer to the symbols $T_i \dots T_j$ inclusive, and $T[i, j)$ to refer to symbols $T_i \dots T_{j-1}$ inclusive. For arrays of phrases, $||$ denotes the append operation : ${\bf b} \gets {\bf b} || a$ appends a variable $a$ to array $\bf b$. Let $p.q$ denote the index, $p.l$ the length, and $p.s$ the innovation of a phrase $p$. Angle brackets denote the tuple of a phrase: $\langle q, l, s\rangle$.

\section{Local Editing Algorithm}
\label{modifying}


\begin{algorithm}[hb!]
	\caption{$\mathit{modify}(i, j, \mathbf{str})$:}
	\label{alg:modify}
\begin{algorithmic}[1]
	\REQUIRE 1. an array of $n'$ phrases, $\mathbf{P}$, which form the LZ-End parsing of the string $\mathbf{T} = T_0, \dots , T_{n-1}$ \\
	2. $0 \leq i \leq j$ \\
	3. a string of symbols $\mathbf{str}$
	\RESULT The array $\mathbf{P}$ which has had symbols $[T_i, T_j)$ removed and $\mathbf{str}$ inserted in their place.
	\STATE $x \gets \mathit{rank}_\text{B}(i)$
	\STATE $l \gets \mathit{rank}_\text{B}(j) - x$
	\STATE $\mathbf{dep} \gets \mathit{find\_dependent\_phrases}(\mathbf{P}, x, x+l$) \hfill (Algorithm~\ref{alg:dependents})
	\STATE $\mathbf{rep}\gets\mathit{find\_replacement\_phrases}( \mathbf{P}$, $i$, $j$, $\mathbf{dep})$ \\ \hfill(\Cref{alg:find_replacement_phrases})
	
	\STATE  $\mathbf{comp\_str} \gets \mathit{encode\_str}(\mathit{P}, i, j, \mathit{str})$ \hfill (Algorithm~\ref{alg:encode_str})
	\STATE delete phrases $\mathbf{P}[\mathit{rank}_B(i), \mathit{rank}_B(j-1)]$
	\STATE insert  $\mathbf{comp\_str}$ into $\mathbf{P}$, between $(\mathit{rank}_B(i) - 1)$ and $\mathit{rank}_B(i)$
	\STATE	$\mathbf{P} \gets \mathit{adjust\_pointers}(\mathbf{P}, x +  \mathbf{comp\_str}.\mathit{size}-1, l + 1,$ $\mathbf{comp\_str}.\mathit{size}, \mathbf{dep}, \mathbf{rep})$ \hfill (Algorithm~\ref{alg:adjust_pointers})
	\RETURN $\mathbf{P}$

\end{algorithmic}
\end{algorithm}

\begin{algorithm}[hb!]
	\caption{$\mathit{find\_dependent\_phrases}(\mathbf{P}, i, j$)}
	\begin{algorithmic}[1]
		\label{alg:dependents}
	\REQUIRE 1. An array $\mathbf{P}$ of LZ-End phrases\\
		2. An integer $i$ denoting the first phrase to be deleted\\
		3. An integer $j$ denoting the first phrase of $\mathbf{P}$ not deleted.
	\RESULT an array of phrase pointers, $\mathbf{dep}$, which point to one or more symbols in $[T_i, T_j)$.

	\STATE $\mathbf{dep} \gets$ an empty array of phrases
	\STATE $\mathbf{dist} \gets$ an array of $\mathbf{P}.\mathit{size} - j$ integers, where $\mathbf{dist}[-1] = 0$.
	\STATE $\mathit{ptr} \gets j$
	\WHILE{$\mathit{ptr} < \mathbf{P}.\mathit{size}$}
		\STATE $\mathbf{dist} [\mathit{ptr} - j] \gets \mathbf{dist}[\mathit{ptr} -  j - 1 ] + \mathbf{P} [\mathit{ptr}].l$
		\IF{$i \leq \mathbf{P}[\mathit{ptr}].q < j \text{ \textbf{or} }\mathbf{dist}[P[\mathit{ptr}].q] \leq \mathbf{P}[\mathit{ptr}].l - 1$}
			\STATE $\mathbf{dep} \gets \mathbf{dep} || \mathit{ptr}$
		\ENDIF
		\STATE $\mathit{ptr} \gets \mathit{ptr} +$ 1.
	\ENDWHILE
	\RETURN $\mathbf{dep}$
	\end{algorithmic}
\end{algorithm} 

This section presents the algorithm to locally edit data compressed using LZ-End parsing. We assume prior knowledge of the index positions of the symbols within $\bf T$ which are to be edited. This is because the shared phrase structure of the LZ77 and LZ-End parsing allows the pattern matching algorithm for LZ77 compressed data~\cite{lz77_pattern_matching} to apply to LZ-End.

We can express any operation we wish to perform on the data ${\bf T} = T_0 \dots T_{n-1}$ as a variation of a generic operation $\mathit{modify}(i, j, \mathbf{str})$ as follows: $i$ indexes the first symbol $\mathbf{T}_i$ to be removed, $j$ the first symbol $\mathbf{T}_j$ after $T_i$ that is not removed, and the symbols of $\mathbf{str}$ replace $[\bf T_i, \bf T_{j})$. Thus, \emph{deletion} happens for $j>i$. For $j = i$, nothing is deleted, and $j < i$ is not allowed. If $\mathbf{str}$ is {\em null}, we have a pure deletion. \emph{Insertion} happens at position $i$. If $j = i$ (no deletion), symbol $i$ moves to the right and we have a pure insertion. If $j>i$ and $\mathbf{str}$ is not {\em null}, we have a \emph{replacement}.

Algorithm~\ref{alg:modify} and its subroutines (\Cref{alg:dependents,alg:find_replacement_phrases,alg:encode_str,alg:adjust_pointers}) together implement the $\mathit{modify}(i$, $j$, $\mathbf{str}$) function. The algorithm first identifies the {\em target} phrases which encode the symbols in $[T_i, T_j)$. It then identifies all dependents of the {\em target} phrases (\Cref{alg:dependents}). Next, the algorithm replaces each {\em dependent} with one or more phrases which do not depend on {\em target} (\Cref{alg:find_replacement_phrases}).
The next step involves appending (prepending) to $\mathit{str}$ the symbols in the first (last) phrase of {\em target} which precede $T_i$ (follow $T_{j-1}$). The updated $\mathit{str}$ is then parsed by the LZ-End algorithm as a separate string. No part of the compressed data is used to encode $\mathit{str}$ (\Cref{alg:encode_str}). The algorithm then replaces the {\em target} phrases by the phrases for $\mathbf{str}$. The final step is to adjust the back-references of each phrase which follows the original {\em target} phrases. This is done to account for the phrases which were inserted and deleted in the previous steps of \Cref{alg:adjust_pointers}. Note that our implementation replaces the inner loop shown here by a binary search, achieving log-linear rather than quadratic running time.

The algorithm to find replacement phrases for each dependent (\Cref{alg:find_replacement_phrases}) first determines the exact symbols $[T_a, T_b]$ in $\mathbf{T}$ which the dependent references, where $a$ and $b$ are set in Steps~\ref{alg:find_replacements:a} and \ref{alg:find_replacements:b}, respectively. The algorithm then checks if the dependent references any symbol(s) preceding the {\em target}. If it does, the dependent's first replacement phrase references the symbols in $[T_a, T_i)$ (Step \ref{alg:find_replacements:prefix}). The same happens in Step \ref{alg:find_replacements:suffix} for symbols $[T_j, T_b]$. The part of the dependent which references one or more phrases in the target is replaced by exact copies of those referenced phrases (Steps \ref{alg:find_replacements:target_begin} -- \ref{alg:find_replacements:target_end}).

\begin{algorithm}[t!]
\caption{$\mathit{find\_ replacement\_ phrases}(\mathbf{P}, i, j, \mathbf{dep}$)}
\label{alg:find_replacement_phrases}
\begin{algorithmic}[1]
	\REQUIRE 1. An array of LZ-End phrases  $\mathbf{P}$ \\
	2. An index $i$ of the first symbol being removed\\
	3. An index $j$ of the first symbol after $i$ not being removed\\
	4. An array $\mathbf{dep}$ of phrases which depend on one or more symbols in  $\mathbf{T}$[$i$, $j)$
	\RESULT a ragged 2-D array of phrases encoding each phrase in $\mathbf{dep}$ without dependence on symbols in $\mathbf{T}[i, j)$
	\STATE $\mathbf{rep} \gets \text{an array of }\mathbf{dep}.size$ many empty arrays of phrases.
	\FOR{$k$ in $[0, \mathbf{dep}.size)$}
		\STATE $b \gets \mathit{select}_B(\mathbf{dep}[k].q)$
		\label{alg:find_replacements:b}
		\STATE $a \gets b - \mathbf{dep}[k].l + 1$
		\label{alg:find_replacements:a}
		\STATE $\mathit{len} \gets \mathbf{dep}[k].l$
		\IF{$a < i$}
		\label{alg:find_replacements:prefix}
			\STATE $\mathbf{rep}$[$k ] \gets \mathbf{rep}$[$k] || \langle\mathit{rank}_B(i)-1, i-a, \mathbf{P}[\mathit{rank}_{B}(i)-1].s\rangle$
			\STATE $\mathit{len} \gets \mathit{len} - i-a$
			\STATE $\mathit{ptr} \gets \mathit{rank}_B(i)$
		\ELSE
			\STATE $\mathit{ptr} \gets \mathit{rank}_B(a)$
		\ENDIF 
		\WHILE{$\mathit{len} >0$ and $\mathit{ptr} \leq \mathit{rank}_B(j$)}
			\label{alg:find_replacements:target_begin}
			\STATE $\mathbf{rep}$[$k] \gets \mathbf{rep}$[$k] || \mathbf{P}[\mathit{ptr}]$
			\STATE $\mathbf{rep}$[$k].l \gets \mathit{Min}(\mathit{len}$, $\mathbf{rep}[k].l$)
			\STATE $\mathit{len} \gets \mathit{len} - \mathbf{P}[\mathit{ptr}].l$
			\STATE $\mathit{ptr} \gets \mathit{ptr} + 1$
		\ENDWHILE
		\label{alg:find_replacements:target_end}
		\IF{$\mathit{len} > 0$}
			\label{alg:find_replacements:suffix}
			\STATE $\mathbf{rep}[k] \gets \mathbf{rep}[k] || \langle \mathbf{dep}[k].q, \mathit{len}, \mathbf{dep}[k].s\rangle$ 
		\ENDIF
	\ENDFOR
	\RETURN $\mathbf{rep}$
\end{algorithmic}
\end{algorithm}

\begin{algorithm}[t!]
\caption{$\mathit{encode\_str}(\mathbf{P}$, $i, j, \mathbf{str}$)}
\label{alg:encode_str}
\begin{algorithmic}[1]
	\REQUIRE 1. an array of LZ-End phrases $\mathbf{P}$\\
	2. An index $i$ of the first symbol $\mathbf{T}[i]$ to be deleted\\
	3. An index $j$ of the first symbol after $\mathbf{T}[i]$ to not be deleted\\
	4. A string of symbols $\mathbf{str}$ to be inserted in place of $\mathbf{T}[i, j)$.
	\RESULT the LZ-End phrases of the symbols in $\mathbf{str}$ as well as the symbols before $i$ which are in the same phrase as $i$, and the symbols after and including $j$ which are in the same phrase as $j-1$.
	\STATE $\mathit{first} \gets \mathit{select}_B(\mathit{rank}_B(i) - 1)  + 1$
	\STATE $\mathit{last} \gets \mathit{select}_B(\mathit{rank}_B(j - 1))$

	\STATE $\mathit{prefix} \gets \mathit{extract}()$\footnotemark[1] the symbols  $\mathbf{T}[\mathit{first}, i)$
	\STATE $\mathit{suffix} \gets \mathit{extract}()$\footnotemark[1] the symbols  $\mathbf{T}[j, \mathit{last}]$
	\STATE $\mathbf{str} \gets \mathit{prefix} || \mathbf{str} || \mathit{suffix}$
	\label{alg:encode_str:extend_str}
	\STATE $\mathbf{comp\_str} \gets$ encode\footnotemark[2] the symbols of $\mathbf{str}$
	\RETURN $\mathbf{comp\_str}$

\end{algorithmic}
\end{algorithm}

\begin{algorithm}[h!]
\caption{$\mathit{adjust\_pointers}(\mathbf{P}, x, l, z, \mathbf{dep}, \mathbf{rep})$}
\label{alg:adjust_pointers}
\begin{algorithmic}[1]
	\REQUIRE 1. An array $\mathbf{P}$ of LZ-End phrases. \\
	2. A pointer $x$ to the first phrase whose index may need adjusting. \\
	3. The number of phrases $l$ which were removed from $\mathbf{P}$. \\
	4. The number of phrases $z$ in the encoding of $\mathbf{str}$.\\
	5. An array $\mathbf{dep}$ of pointers to phrases which depended on phrases which have been removed. \\
	6. A ragged 2-D array $\mathbf{rep}$ of phrases which were inserted in place of each phrase in $\mathbf{dep}$.
	\RESULT The array $\mathbf{P}$, once its index pointers have been corrected to account for the modification.
	\FOR{$i$ in $[x, \mathbf{P}.\mathit{size})$}
	\label{alg:adjust_indexes_start}
		\IF{$P[i].q \geq x$}
			\STATE $P[i].q \gets P[i].q  + z - l$
			\FOR{$j$ in $[0, \mathbf{dep}.\mathit{size})$}
				\IF{$P[i].q \geq \mathbf{dep}[j]$}
					\STATE $P[i].q \gets P[i].q + \mathbf{rep}[j].\mathit{size}$
				\ELSE
					\STATE break
				\ENDIF
			\ENDFOR
		\ENDIF
	\ENDFOR
	\RETURN $\mathbf{P}$
	\label{alg:adjust_indexes_end}
\end{algorithmic}
\end{algorithm}

\footnotetext[1]{Using Kreft and Navarro's $\mathit{extract}()$ algorithm \cite{lzend}}
\footnotetext[2]{Using the LZ-End parsing algorithm \cite{lzend}}

Note that replacement does not change existing phrase boundaries, i.e., any phrases referencing a dependent can instead reference the corresponding final replacement phrase. The worst case run-times of the subroutines are as listed, where the variables are as defined in \Cref{alg:modify}. The derivation of the runtime complexity for each subroutine is found in the Appendix:
\begin{itemize}
	\item Find dependent phrases (\Cref{alg:dependents}): $\Theta(n' - \mathit{rank}_{B}(j))$.
	\item Find replacement phrases (\Cref{alg:find_replacement_phrases}): $O((n' - \mathit{rank}_B(j)) $ $\times (\mathit{rank}_B(j) - \mathit{rank}_B(i) + 2))$.
	\item Encoding $\mathbf{str}$ (\Cref{alg:encode_str}): $O(S \log s)$, $S \leq sl$, where $l$ is the length of the longest phrase encoding $\mathbf{str}$, and $s$ is the size of $\mathbf{str}$ after Step~\ref{alg:encode_str:extend_str} of~\Cref{alg:encode_str}.
	\item Adjust pointers (\Cref{alg:adjust_pointers}): $O((n' - \mathit{rank}_B(j)) \times \log(n' - \mathit{rank}_B(j)))$. This occurs when every phrase after $\mathit{rank}_{B}(j)$ was a dependent of all preceding phrases up to and including $\mathit{rank}_{B}(j)$ before modification. Each dependent is replaced by at least one phrase which references phrases preceding $\mathit{rank}_{B}(i)$ (i.e., needs no index pointer adjustment) and at most by one phrase with index $> \mathit{rank}_{B}(j)$ (which needs adjustment).
\end{itemize}

\section{Evaluation}

The LZ-End parsing is known to be \textit{coarsely optimal}~\cite{lzend_coarsely_optimal}. That is, for any $k$, the compression ratio is asymptotically bounded by the $k$-th order entropy $H_k$ of the string. The proof of coarse optimality depends on the assumption that each phrase in the LZ-End parsing of a string represents a unique sequence of symbols. Although this assumption is valid for the LZ-End parsing, the $\mathit{modify}()$ operation does not preserve phrase uniqueness. The coarse optimality of the LZ-End parsing therefore does not hold after the application of the $\mathit{modify}()$ operation. This section empirically evaluates the effect of the $\mathit{modify}()$ operation on the compressed data.

\subsection{Metrics}

The compression performance of the LZ-End algorithm has already been extensively evaluated theoretically and empirically \cite{lzend}. We therefore focus on evaluating the effect which the $\mathit{modify}()$ operation has on the compressed data. The evaluation considers the following factors:
\begin{itemize}
	\item the type of modification -- that is, insertions, deletions, and replacements.
	\item the size of the modification (as a fraction of overall file size).
	\item the effect of successive modifications.
	\item the characteristics of the file being modified.
\end{itemize}

To this end, the experiments consisted of applying the same set of modifications to two different sets of files. The first set was the Canterbury Corpus \cite{canterbury_corpus}, which is well-known in literature. The second part of the evaluation consisted of generating a set of files of calibrated entropy, as described by Ebeling \textit{et al.} \cite{titchener_calibrated_entropy}. These two sets of experiments allow the evaluation of the $\mathit{modify}()$ function on a set of well-known and easily accessed files, as well as extrapolation of the results to files of various entropy.

The effect of $\mathit{modify}()$ is measured by the modification ratio (MR). This is calculated as the size of the compressed data following the application of $\mathit{modify}()$, divided by the size of the compressed data if it had been decompressed, modified and recompressed.

The following set of modifications were carried out on the files:
\begin{itemize}
	\item Incremental modifications (\Cref{fig:incremental_evaluation} and left column of \Cref{table:canterbury}).

	The ratio reported is the mean MR of each file following 100 insertions, 100 deletions, or 100 replacements. Each modification was 0.5\% of the original file size, so that each set of insertions, deletions, and replacements affected 50\% of the original file.

	\item Size of the modification (\Cref{fig:sizes_evaluation} and middle columns of \Cref{table:canterbury}).

	The reported MR is the mean MR of insertions, deletions and replacements of various sizes between 5\% and 95\% of the file size.

	\item Position of the modification (\Cref{fig:positions_evaluation} and rightmost columns of \Cref{table:canterbury}).

	The reported MR is the mean MR after a modification of size 0.5\% of the file size, made at various points between 0.05 and 0.95 of the file length. For a file of 100 bytes, an edit made at 0.05 starts at the 5\textsuperscript{th} byte, while an edit made at 0.95 starts at the 95\textsuperscript{th} byte.
	
\end{itemize}

Although not particularly time efficient, this algorithm represents the first step in a new and interesting direction to improve the efficiency of compression functions in the future.

\subsection{Implementation Notes}

The $\mathit{modify}()$ operation does not require any change to the LZ-End parsing or decoding algorithms. The phrases following an application of $\mathit{modify}()$ are valid LZ-End phrases (except that the phrases are no longer unique). This means that the decoder does not need to be changed at all. 

The implementation of the LZ-End parser did not optimise the encoding for short phrases. Rather, the implementation encodes even single characters in its own phrase. This is similar to the original LZ77 parsing \cite{lz77}, and is unlike the more optimised version of Strorer and Szymanski \cite{lzss}. Whether or not fast local decoding of LZ-End data is still possible under an LZSS-type parsing applied to the LZ-End algorithm has not been investigated. 

The calibrated entropy strings were produced by a bipartition $c \in [0, 1]$ of the logistic map, as proposed by Ebeling et al.\ in~\cite{titchener_calibrated_entropy}. 
To further demonstrate dependence on entropy, one can apply an FIR filter to the bit streams generated by the logistic map to correlate subsequent bits and obtain a lower entropy string.
$F_n$, the $n$-th output value produced by the FIR filter, results from the logistic map bit sequence $S_n$ via an equation such as the following:
\begin{equation*}
	F_{n} = 0.5 \times S_{n} + 0.1 \times S_{n - 1} + \dots + 0.1 \times S_{n-5}\nonumber \text{\qquad where } S_{-1} = \dots = S_{-5} = 1
\end{equation*}

The value of the $n$-th bit is then set using a bipartition at 0.4, similar to that of the Logistic Map. This means that the FIR filter string will have a bit value of 1 wherever the logistic map string has a bit value of 1 or wherever four or more of the preceding five bits in the logistic map string are set to 1.
Note that the order and the coefficients of the FIR filter above are chosen somewhat arbitrarily -- all we need is \emph{some} correlation. This can be achieved with a large range of FIR designs.

Ten strings of 50000 bytes each were generated for each value of the noise amplitude $\xi$ in Table~\ref{table:calibration_values} with calibration $c=0.5$~\cite{titchener_calibrated_entropy}. The table shows the average, minimum and maximum size in KiB to which these strings compressed. At low entropies, FIR filtering has little impact on the compressed size, but as expected reduces the compressed size somewhat for larger entropies. This shows that the effects seen cannot be mere artefacts of the logistic map generation.

Each experiment was run with three kinds of $\mathbf{str}$:
\begin{itemize}
	\item a low-entropy $\mathbf{str}$ consisting only of a series of ASCII symbols ``a'',
	\item a medium-entropy $\mathbf{str}$ which compresses to half its size, and
	\item a high-entropy $\mathbf{str}$ which doesn't compress.
\end{itemize}  
The mean MR for these three values of $\mathbf{str}$ is reported.

We have conducted extensive validation to confirm correctness of $\mathit{modify}()$. This involved incrementally making a series of 100 random modifications of random sizes and in random positions within each file of the Canterbury Corpus \cite{canterbury_corpus}, and confirming that each file decompressed correctly.

\subsection{Results}

\begin{table}[t]
\setlength{\tabcolsep}{14pt}
	\centering
	\caption{Compressed sizes of strings with calibrated entropy}
	\begin{tabular}{@{}crrr|rrr@{}}
		\toprule
		& \multicolumn{3}{c|}{Logistic map} & \multicolumn{3}{c}{Logistic map + FIR filter}\\
		$\xi$ & avg. & min. & max. & avg. & min. & max. \\
		\cmidrule{1-7}
		0.0001 & 8.88 & 8.79 & 8.94 & 8.88 & 8.82 & 8.94 \\
		0.00025 & 11.91 & 11.69 & 12.71 & 11.91 & 11.69 & 12.74 \\
		0.0005 & 15.45 & 15.38 & 15.50 & 15.45 & 15.42 & 15.48 \\
		0.00075 & 17.40 & 17.31 & 17.46 & 17.41 & 17.31 & 17.47 \\
		0.001 & 18.95 & 18.90 & 19.03 & 18.94 & 18.86 & 19.01 \\
		0.0025 & 26.40 & 26.26 & 26.51 & 26.86 & 26.68 & 26.98 \\
		0.005 & 32.12 & 31.98 & 32.22 & 32.81 & 32.73 & 32.92 \\
		0.0075 & 35.88 & 35.77 & 35.97 & 35.03 & 34.92 & 35.15 \\
		0.01 & 38.94 & 38.82 & 39.05 & 35.88 & 35.73 & 36.01 \\
		0.025 & 52.60 & 52.49 & 52.78 & 38.16 & 38.06 & 38.28 \\
		\bottomrule
	\end{tabular}
	\label{table:calibration_values}
\end{table}

\begin{table}[t]
	\centering
	\caption{Modification ratios of the Canterbury Corpus}
	\label{table:canterbury}
	\begin{tabular}{|c|c|ccc|ccc|}
		\hline
		\textbf{File:} & Incremental MR & \multicolumn{3}{c|}{Sizes} & \multicolumn{3}{c|}{Positions}\\
		& & 0.05 & 0.5 & 0.95 & 0.05 & 0.5 & 0.95\\ 
		\hline
		
		aaa.txt & 242.0 & 1.565 & 1.527 & 1.590 & 1.889 & 1.718 & 1.650 \\ 
		alice29.txt & 1.169 & 1.015 & 1.155 & 1.527 & 1.014 & 1.002 & 1.000 \\ 
		alphabet.txt & 32.20 & 1.557 & 1.481 & 1.471 & 1.812 & 1.809 & 1.725 \\ 
		asyoulik.txt & 1.217 & 1.012 & 1.134 & 1.511 & 1.012 & 1.003 & 1.000 \\ 
		bible.txt & 1.518 & 1.018 & 1.181 & 1.567 & 1.014 & 1.004 & 1.045 \\ 
		cp.html & 1.394 & 1.011 & 1.115 & 1.273 & 1.015 & 1.002 & 1.004 \\ 
		E.coli & 1.335 & 1.010 & 1.134 & 1.431 & 1.059 & 1.002 & 1.000 \\ 
		fields.c & 1.597 & 1.025 & 1.091 & 1.216 & 1.013 & 1.006 & 1.004 \\ 
		grammar.lsp & 1.747 & 1.037 & 1.187 & 1.140 & 1.024 & 1.032 & 1.004 \\ 
		kennedy.xls & 1.205 & 1.014 & 1.070 & 1.318 & 1.006 & 1.001 & 1.000 \\ 
		lcet10.txt & 1.701 & 1.013 & 1.142 & 1.272 & 1.015 & 1.003 & 1.000 \\ 
		pi.txt & 1.275 & 1.010 & 1.121 & 1.350 & 1.011 & 1.002 & 1.000 \\ 
		plrabn12.txt & 1.302 & 1.011 & 1.141 & 1.375 & 1.012 & 1.002 & 1.000 \\ 
		ptt5 & 1.577 & 1.008 & 1.291 & 1.256 & 1.001 & 1.003 & 1.001 \\ 
		random.txt & 1.167 & 1.010 & 1.118 & 1.183 & 1.011 & 1.002 & 1.000 \\ 
		sum & 1.406 & 1.003 & 1.436 & 1.468 & 1.013 & 1.001 & 1.009 \\ 
		world192.txt & 1.326 & 1.011 & 1.132 & 1.239 & 1.016 & 1.002 & 1.002 \\ 
		xargs.1 & 1.479 & 1.017 & 1.145 & 1.118 & 1.015 & 1.008 & 1.005 \\ \hline
	\end{tabular}
\end{table}

\begin{figure*}[hp]
	\centering
	\begin{subfigure}{\textwidth}
		\centering
		\includegraphics[width=10cm, trim=2.85cm 20cm 6cm 3cm, clip]{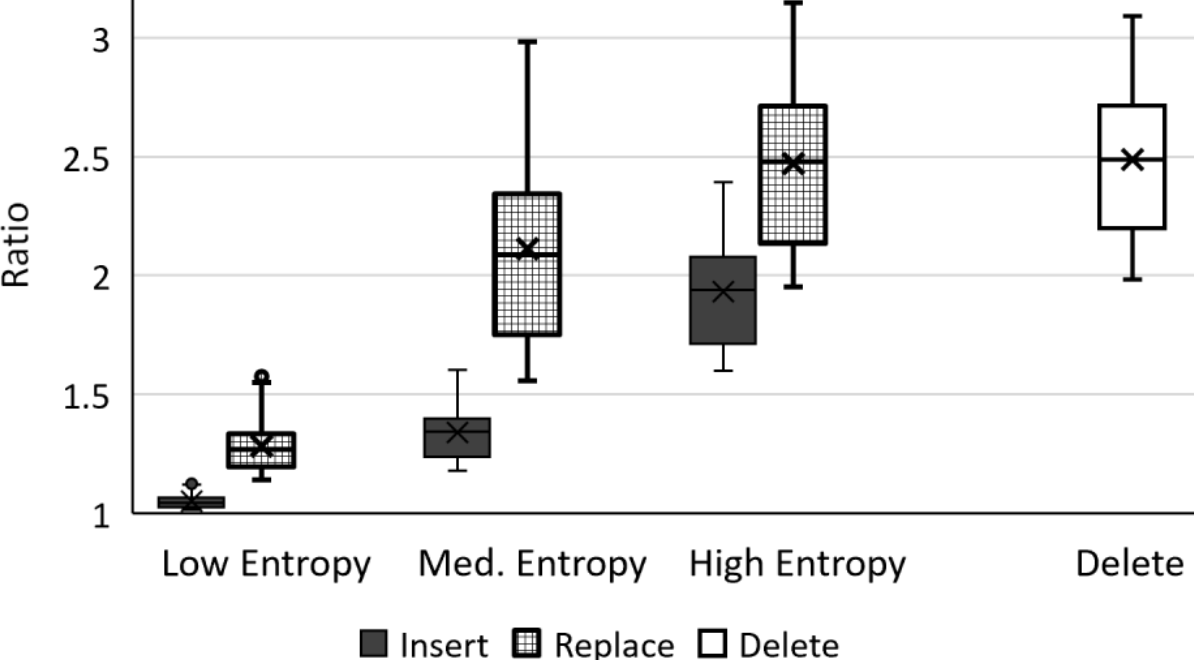}
		\caption{Effect of 100 incremental modifications on the compressed size of	strings with calibrated entropies. The High, Medium and Low Entropy classes refer to the entropy of the $\mathbf{str}$ which is inserted.\vspace{0.24in}}
		\label{fig:incremental_evaluation}
	\end{subfigure}
	\begin{subfigure}{\textwidth}
		\centering
		\includegraphics[width=10cm, trim=3cm 3.15cm 6cm 19.6cm, clip]{CalibratedResults.pdf}
		\caption{MRs vs.\ relative size of a single modification. The size of the insertion does not affect the MR, while the deletions have a larger effect than replacements. \vspace{0.24in}}
		\label{fig:sizes_evaluation}
	\end{subfigure}
	
	\begin{subfigure}{\textwidth}
		\centering
		\includegraphics[width=10cm, trim=3cm 11.8cm 6cm 10.95cm, clip]{CalibratedResults.pdf}
		\caption{Effect of modification position of a single local modification on the MR. The insertion ratio is independent of position, while deletion and replacement are nearly identical for these small modifications.}
		\label{fig:positions_evaluation}
	\end{subfigure}
	\caption{Evaluation of various modifications carried out on the strings generated using the logistic map and the FIR filter.}
	\label{fig:calibrated_entropy_experiments}
\end{figure*}

The first column in \Cref{table:canterbury} shows the effect of 100 incremental modifications. The values reported are the mean ratio for insertions, deletions and replacements, for high, medium and low entropy $\mathbf{str}$ sources. \Cref{fig:incremental_evaluation} shows the breakdown of how these operations and the entropy of the $\mathbf{str}$ relate to one another. As expected, modification most negatively affects the lowest entropy files (e.g., {\tt aaa.txt} or {\tt alphabet.txt}). This is because a single phrase of data from a low-entropy source encodes much more data than for a higher entropy source. We see from \Cref{fig:incremental_evaluation} that the insertion of a low entropy $\mathbf{str}$ has a much smaller effect than for a $\mathbf{str}$ of higher entropy. We also see that insertion has less of an effect than the replacement and deletion operations. This is because the degradation of the compression ratio occurs mainly in the replacement of dependent phrases. An insertion will have few dependent phrases, since no phrases are being removed. The replacement operation can be viewed as a deletion followed by an insertion. This explains the greater degradation in the MR which occurs as a result of replacements compared to insertions.

The middle columns of \Cref{table:canterbury} shows the effect of the size of the edit on the MR. \Cref{fig:sizes_evaluation} demonstrates how the deletion, insertion and replacement operations compare for different size edits. Insertion has a constant effect on the MR, whereby the MR stays just above 1 for any size modification. Deletion has the highest MR for all sizes of edit, with the MR for replacements sandwiched between that of deletion and insertion. The MR of a deletion and replacement generally increases for larger edits. However, the experiments on the Canterbury Corpus show that this is not a universal rule, and depends on the characteristics of the individual file.

Finally, the rightmost colums of \Cref{table:canterbury} shows the effect of the position of the edit on the MR. \Cref{fig:positions_evaluation} shows how the insertions, deletions and replacements relate to one another for the strings of calibrated entropy. Generally, insertions have constant effect, while replacements and deletions have almost identical effect. Deletions and replacements at the start of a file have the greatest effect on MR, and this quickly tapers off. This is due to the fact that the initial phrases at the very start of a file have many dependencies. However, specific files such as {\tt bible.txt} demonstrate exceptions to this principle. The edits made in these tests are very small (0.5\% of the file size), hence the small effect on MR. Once again, files of low entropy ({\tt aaa.txt} and {\tt alphabet.txt}) have the greatest MR.

\section{Conclusion and Future Work}

This paper presented the first algorithm to locally modify data compressed by a type of Lempel-Ziv parsing. Existing results on the LZ-End algorithm allowed data to be found and decoded without decompression. Modifying the compressed data, however, was previously not possible without decompressing the entire compressed string first. Our implementation of the algorithm was subjected to extensive testing to ensure correctness.

Investigations using Canterbury Corpus data and files of calibrated entropy show that small local modifications usually have a very small effect on the compression ratio of the data, as do insertions of any size and position, compared to what would be the case when decompressing, modifying and recompressing the data. Local deletions and replacements which affect a larger part of the data have a greater effect on the compression. The effect which the position of the local modification has on the compression is dependent on the properties of the data itself; in the general case, however, modifications made towards the start of the data have a larger effect on the compression ratio.

The algorithm to modify the data is not particularly time-efficient, as it is the first of its kind for a LZ-type parser. The primary factor limiting the performance of the algorithm is the lack of a sliding window in the LZ-End parser. The LZ77 algorithm does incorporate a sliding window into its parsing~\cite{lz77}; however, the LZ77 parser does not currently support local decoding. Should the LZ77 parsing allow local decoding, the methods presented in this paper can be used to allow random editing as well. This random editing will also be faster than the current method for LZ-End, as the number of dependent phrases will be strictly limited by the size of the sliding window.

\section*{Acknowledgment}

The authors would like to thank Mark Titchener for his suggestions regarding generating the calibrated entropy strings.

\newpage

\appendix

\subsection{Complexity of \Cref{alg:dependents}}

\Cref{alg:dependents} is a fairly straightforward linear algorithm. It iteratively checks each phrase which follows the modification location, to determine whether that phrase is dependent on the section of data being edited. This iterates from the phrase containing the first character which is not removed ($\mathit{rank}_B(j)$) until the end of the compressed data ($n'$).

The complexity of Algorithm2 is therefore $\Theta(n' - \mathit{rank}_B(j))$.

\subsection{Complexity of \Cref{alg:find_replacement_phrases}}

The $\mathit{rank}_B()$ and $\mathit{select}_B()$ operations are constant-time array look-ups \cite{lz77}. We therefore focus on the {\tt for} and {\tt while} loops in the algorithm.

The {\tt for} loop iterates once for each dependent phrase and the {\tt while} loop iterates once for each phrase which is referenced by that dependent. The worst-case scenario is when all phrases following the location of the edit are dependent, and when each dependent phrase references all phrases which are being edited.

The number of phrases which follow the edit location is $n' - \mathit{rank}_B(j)$, and the number of phrases which are edited is $\mathit{rank}_B(j) - \mathit{rank}_B(i) + 2$.

The overall complexity of the $\mathit{find\_replacement\_phrases()}$ subroutine is therefore $O((n' - \mathit{rank}_B(j)) \times (\mathit{rank}_B(j) - \mathit{rank}_B(i) + 2))$.

\subsection{Complexity of Algorithm 4}

\Cref{alg:encode_str} simply encodes the symbols of $\mathbf{str}$ after it has been extended in the first part of the subroutine. The extended $\mathbf{str}$ is then encoded as its own string, without any backreferences to previous parts of the compressed data. The complexity of calculating the LZ-End parsing of a string is $O(S \log s)$, where $S \leq sl$, $l$ is the length of the longest phrase and $s$ is the length of the string \cite{lzend}.

\subsection{Complexity of Algorithm 5}

The final subroutine iterates through every phrase which follows the point at which the edit was made, and adjusts the back-references of each phrase which points to some phrase following the edit location. 

Each dependent phrase is replaced by two or more replacement phrases in \Cref{alg:find_replacement_phrases}. The replacement phrases each point to phrases preceding the edit location, and these back-references do not need to be adjusted. The only exception is the final replacement phrase for each dependent phrase, which may point to a phrase immediately after the edit location. The worst-case scenario is when each phrase which comes after the edit location is dependent on the edited phrases (and its backreference therefore needs to be adjusted).

The inner {\tt for} loop in Step 4 of the algorithm is for didactic purposes only. In practice, a binary search is used, achieving logarithmic search time.

Pior to making the modification, there were $n' - \mathit{rank}_B(j)$ many phrases after the edited \textit{target} phrases. These each therefore have at most one replacement phrase which satisfies the criteria in Step 2 of the algorithm, whereby its backreference needs adjusting. 

The complexity of this subroutine is therefore $O((n' - \mathit{rank}_B(j)) \times \log(n' - \mathit{rank}_B(j)))$. 
\end{document}